\documentclass[aps,prd,twocolumn,nofootinbib,floatfix]{revtex4-1}
\usepackage{setspace}
\usepackage{amsthm}
\theoremstyle{definition}

\usepackage{bbold}
\usepackage{graphicx}
\usepackage{dcolumn}
\usepackage{multirow}
\usepackage{subfigure}
\usepackage{times,mathptm}
\usepackage{float}
\usepackage{color}
\usepackage{amsmath,amsfonts}
\usepackage{mathptmx}
\usepackage{mathrsfs}
\usepackage{bbm}
\usepackage{bm}
\usepackage{xfrac}

\newcommand{\beq}{\begin{equation}}
\newcommand{\eeq}{\end{equation}} 
\newcommand{\bea}{\begin{eqnarray}}
\newcommand{\eea}{\end{eqnarray}}

\newcommand{\E}{\mathcal{E}_0}
\newcommand{\V}{{\cal V}}

\newcommand{\Sc}{S$_\text{c}$\ }

\renewcommand{\d}{\delta}

\renewcommand{\l}{\lambda}

\newcommand{\ophi}{\overline{\phi}}

\renewcommand{\b}{\beta}
\renewcommand{\a}{\alpha}

\newcommand{\tr}{\text{Tr}}

\newcommand{\vx}{{\vec{x}}}
\newcommand{\vy}{{\vec{y}}}
\newcommand{\vz}{\vec{z}}

\newcommand{\m}{\mu}
\newcommand{\pbar}{\overline{\psi}}

\newcommand{\g}{\gamma}

\newcommand{\s}{\sigma}
\renewcommand{\k}{\kappa}

\renewcommand{\th}{\theta}

\newcommand{\oh}{\frac{1}{2}}

\newcommand{\dg}{\dagger}
\newcommand{\non}{\nonumber}
\renewcommand{\t}{\tau}
\newcommand{\rf}[1]{(\ref{#1})}
\newcommand{\ra}{\rightarrow}
\newcommand{\pa}{\partial}
\renewcommand{\vec}[1]{\bm #1}

\usepackage{ulem}

\begin{document}

\title{Aspects of the Higgs phase in SU(2)$\times$U(1) lattice gauge Higgs theory} 

\bigskip
\bigskip

\author{Shivam Gangwani, Jeff Greensite, and Anass Oualla}
\affiliation{Physics and Astronomy Department \\ San Francisco State
University  \\ San Francisco, CA~94132, USA}
\bigskip
\date{\today}
\vspace{60pt}
\begin{abstract}

\singlespacing
 
 Using a simplified lattice version of the electroweak sector of the standard model, with
 dynamical fermions excluded, we determine at fixed Weinberg angle the transition line between the confined phase
 and the Higgs phase, the latter defined as the region where the global center subgroup of the gauge group 
 is spontaneously broken, and ``separation of charge'' confinement disappears. We then
 search, via lattice Monte Carlo simulations, for possible neutral vector bosons in the Higgs
 region, apart from the photon and Z.  There are numerical indications of 
 a ``light Z''  in the lattice data (along with the photon and the Z), but a lack of the expected scaling of the light
 mass particle excludes any firm conclusions about the physical spectrum.

 \end{abstract}

%
%
%
\maketitle
 
\singlespacing

\section{\label{Intro} Introduction}

   The location of the Higgs phase of the Standard Model in the space of couplings depends on what one means by the Higgs phase.  The earliest numerical work on the zero temperature phase diagram that we are aware of was by Shrock in 
\cite{*Shrock:1985un,*Shrock:1985ur,*Shrock:1986av}, and a much later treatment was by Veselov and Zubkov in \cite{Zubkov:2008gi,Zubkov:2009bk}. Of course there have been a great many lattice treatments of the electroweak phase transition at finite temperature (mostly in the SU(2) gauge Higgs model, e.g.\ \cite{Fodor:1994sj}), as well as other topics in the electroweak theory (such as vacuum stability \cite{Kuti:2008PT}), but this article is concerned with phase structure at zero temperature. If the Higgs and confinement phases of the electroweak sector were entirely separated by a boundary of thermodynamic transition, then determination of the phase
diagram at zero temperature would be fairly straightforward.  However, as we know from the work of  
\cite{Osterwalder:1977pc,Fradkin:1978dv,Banks:1979fi}, this is generally not the case
in gauge Higgs theories with the Higgs field transforming in the fundamental representation of the gauge group.  In the absence
of such a boundary, it is necessary to carefully define what is meant by the ``spontaneous breaking'' of a gauge symmetry, which for
local symmetries is actually ruled out by the Elitzur theorem \cite{Elitzur:1975im}.  Our view, advocated in \cite{Greensite:2020nhg}, is that the Higgs phase is distinguished from the confinement and massless phases of a gauge Higgs theory by the
spontaneous breaking of the global center subgroup (GCS) of the gauge group, e.g.\ the spontaneous breaking of the global $Z_N$ subgroup of a local SU(N) gauge group, and this global subgroup transforms the matter fields of the theory, but does not
affect the gauge fields.  While the breaking of this symmetry may or may not be accompanied by a thermodynamic transition, it nonetheless is accompanied by physical effects, in particular the disappearance of metastable flux tube states in the Higgs phase, and corresponding
absence of linear Regge trajectories.

    In this article we first map out the transition line, using the above symmetry breaking criterion, in a simplified SU(2)$\times$U(1) lattice gauge Higgs theory, with a unimodular Higgs
 field (corresponding to infinite Higgs mass), and no dynamical fermions.  Keeping the Weinberg angle fixed, the Wilson coupling
 for the SU(2) lattice field and a single parameter in the Higgs sector define a two-dimensional parameter space, and  the transition
 line we compute lies in this plane.
 
     A second objective is a search, via lattice Monte Carlo simulations, for new vector boson particle states within the Higgs phase whose
 boundary we have located.
 Here we note that physical particles in the electroweak theory are, in some sense, composite objects, and in any quantum theory
 composite objects generally have a spectrum of excitations.  As 't Hooft emphasized many years ago \cite{tHooft:1979yoe}, weak isospin is actually ``confined'' in the electroweak sector.  Certainly there is a distinction between electrons and neutrinos, but the physical particles
 are actually composites, with weak isospin screened by the Higgs field.  The same point was made by Frohlich et al.\  \cite{Frohlich:1981yi} and  by Banks and Rabinovici \cite{Banks:1979fi}.  So, non-perturbatively, 
 excited fermion states might exist. Simulations in SU(3) gauge Higgs theory strongly suggest that this may be the case \cite{Greensite:2020lmh}.  Unfortunately we have no reliable lattice formulation of a chiral gauge theory, so the existence
 of such excitations in the electroweak theory is not amenable to numerical investigation.  But there is no such obstruction to a search for new states with the quantum numbers of the photon and the Z boson in a lattice formulation, providing dynamical fermions are excluded.  We will see that there is evidence,
 in the simplified theory, for a new vector boson state, with a mass significantly lighter than the Z boson, but unfortunately that
 mass does not scale with lattice couplings in the way one would expect.  This prevents us from drawing strong conclusions about the
 existence of such particles in the actual electroweak sector.
 
     In section \ref{theory} we review for completeness some of the ideas presented in \cite{Greensite:2020nhg} regarding the Higgs phase as a phase of broken
 global center symmetry.  These ideas are extended to SU(2)$\times$U(1) gauge Higgs theory in section \ref{line}, where the confinement to Higgs transition line, at fixed Weinberg angle, is determined.  Section \ref{newZ} describes the results of our search for
new Z boson-like excitations, and the last section contains our conclusions.

\section{\label{theory} The symmetric/Higgs phase distinction}

    Our view, as just mentioned, is that Higgs phase is distinguished from the confinement and Coulomb phases of a gauge Higgs theory by the spontaneous breaking of the global center subgroup (GCS) of the gauge group, and this breaking is accompanied by physical effects.  In the case of a massless to Higgs transition, this is simply the appearance of a mass gap.  The effect in the confinement to Higgs transition
is more subtle; it is the loss of metastable color electric flux tubes which would be associated with linear Regge trajectories.  This is a transition between confinement types, which were termed, in \cite{Greensite:2020nhg}, ``separation of charge'' (S$_\text{c})$ confinement and color (C) confinement, and it is not necessarily accompanied by a thermodynamic phase transition.  The GCS is a subgroup of the gauge group which transforms matter fields but not gauge fields, and should not be confused with a different center symmetry which transforms gauge
fields but not matter fields, with Polyakov lines as an order parameter for symmetry breaking.\footnote{In the case of SU(2) gauge Higgs theory there is a larger global SU(2) symmetry, generally known as ``custodial'' symmetry, which transforms the Higgs but not the
gauge fields, and it was the SU(2) gauge group that was mainly considered in \cite{Greensite:2020nhg}.  But SU(2) is a special case, and the relevant symmetry in the general case is the global center subgroup of the gauge group.}

    We define a ``charged'' state to be a physical state, satisfying the Gauss Law constraint, which transforms covariantly under an unbroken GCS.  The simplest illustration of a charged state in an infinite volume, and the motivation for this definition, is a state containing a single static fermion coupled to the quantized Maxwell field (no dynamical matter fields), presented long ago by Dirac \cite{Dirac:1955uv}.  In $A_0=0$ gauge the charged state is
\beq
  \Psi_{\text{chrg}} =  \pbar(x) \rho(x;A) \Psi_0  \ ,
 \eeq
 where $\Psi_0$ is the ground state, $\pbar$ creates a static fermion, with
 \bea
       \rho(x;A) &=&  \exp\left[-i {e\over 4\pi} \int d^3z ~ A_i(\vz) {\pa \over \pa z_i}  {1\over |\vx-\vz|}  \right]
 \label{rho}
 \eea
 and \cite{Misner:1973prb}
 \beq
        \Psi_0[A] =  \exp\left[ - \int d^3x \int d^3y ~  
         { \nabla \times {\bf A}(x) \cdot \nabla \times {\bf A}(y) \over 16\pi^3 |x-y|^2} \right] \ .
\label{MTW}
\eeq
 Let $g(x) = e^{i\th(x)}$ be an arbitrary U(1) gauge transformation.  The ground state is obviously invariant under this transformation,
 while $\pbar(x) \ra e^{-i\th(x)} \pbar(x)$ transforms covariantly.  The field $\rho(x)$, however, is almost but not quite covariant under the
 gauge transformation.  Let $\th_0$ be the zero mode of $\th(x)$, i.e.\ $\th(x) = \th_0 + \tilde{\th}(x)$.  Then it is easy to see that
 \beq
            \rho[x;A] \ra e^{i\tilde{\th}(x)} \rho[x;A] \ ,
\eeq
and therefore, under an arbitary gauge transformation
\beq
          \Psi_{\text{chrg}} \ra e^{-i\th_0}  \Psi_{\text{chrg}} \ .
\eeq
In other words, a charged state in U(1) gauge theory is almost but not entirely gauge invariant.  It transforms covariantly
under the global center subgroup of the gauge group, consisting of transformations $g(x) = e^{i\th_0}$.  If there is a dynamical matter (e.g.\ scalar) field $\phi$ in the theory which transforms in the same way as the static fermion, then one can also construct neutral states containing the static fermion, such as
\beq
          \Psi_{\text{neutral}} = \pbar(x) \phi(x) \Psi_0  \  ,
\eeq
which is invariant under the full gauge group.  However, this sharp distinction between charged and neutral states breaks down if, upon inclusion of dynamical matter, the GCS is spontaneously broken, as in the Higgs phase of the abelian Higgs model.  In situations of that kind, states which transform covariantly under the GCS are not necessarily orthogonal to neutral
states such as $\Psi_{\text{neutral}}$ above. 

      All of this extends directly to non-abelian gauge Higgs theories.  The operator $\rho(x;A)$ is one example of what has been called a
 ``pseudomatter'' operator \cite{Greensite:2020nhg}, defined to be a functional of the gauge field which transforms like a matter field in the fundamental representation of the gauge group, except under gauge transformations belonging to the global center subgroup of the gauge group.  Examples in non-abelian lattice gauge theories include gauge transformations to physical gauges, and eigenstates $\xi_n^a(x;U)$ of the covariant lattice Laplacian on a time slice
 \beq
            -D^{ab}_{xy}[U] \xi^b(y;U) = \l_n \xi_n^a(x;U) \ , 
 \eeq
 where
 \bea
    D^{ab}_{\vx \vy} &=& \sum_{k=1}^3 \left[2 \d^{ab} \d_{\vx \vy} - U_k^{ab}(\vx) \d_{\vy,\vx+\hat{k}}  - U_k^{\dg ab}(\vx-\hat{k}) \d_{\vy,\vx-\hat{k}}   \right]  \non \\
\label{Laplacian}
\eea
is the lattice Laplacian, and superscripts are color indices.  Note that since the lattice gauge field $U_\m(x)$ is unaffected by the global center subgroup of the gauge group, so are the $\xi_n$.  Using the pseudomatter operators $\xi_n$, or any other pseudomatter operators, we can construct physical states in gauge Higgs theories such as
\beq
  \Psi_{\text{chrg}} =  \pbar^a(x) \xi^a(x;U) \Psi_0 [U,\phi]  \ ,
\eeq
which transform covariantly under the GCS, and are charged providing that this symmetry is not spontaneously broken.  A gauge Higgs
theory, or any gauge theory with matter in the fundamental representation of the gauge group, is in the separation of charge (\Sc) confining phase if the energy expectation value of any state of this kind, above the vacuum energy, is infinite.  In a finite volume one can construct
finite energy states by creating separated charges
\beq
\Psi  = \pbar^a(x) \xi^a(x;U) \xi^{b\dg}(y;U) \psi^b(y) \Psi_0 [U,\phi]  \ .
\eeq
In the \Sc  phase the energy of such states tends to infinity as $|x-y| \ra \infty$ for {\it any} pseudomatter operator $\xi$.  In fact the
statement holds for any state of this form
\beq
\Psi  = \pbar^a(x)  V^{ab}(x,y;U) \psi^b(y) \Psi_0 [U,\phi]  \ ,
\eeq
where $ V^{ab}(x,y;U)$ is any functional of the gauge field only, transforming bicovariantly at the points $x,y$.
\Sc  confinement exists only when the GCS symmetry is unbroken. In the Higgs phase, where the GCS symmetry is broken, it can
be shown that there are always some choices for $V(x,y;U)$ which result in finite energy states as $|x-y|\ra \infty$ \cite{Greensite:2020nhg}.
The question is how to construct a gauge-invariant order parameter for the symmetry breaking.

   Of course it is nonsense to regard $\langle \phi \rangle$ as an order parameter.  In the absence of gauge fixing this quantity is zero regardless of the couplings; in a unitary gauge it is non-zero regardless of the couplings, and in other gauges it may be zero or non-zero
in various regions of coupling constant space, depending on the choice of gauge \cite{Caudy:2007sf}.  Following  \cite{Greensite:2020nhg}, we construct an order parameter starting from
\bea
                 e^{-H[U,\phi]/kT} &=& \int DU_0(DU_k D\phi]_{t\ne 0} e^{-S} \non \\
                                            &=&  \sum_n |\Psi_n[U,\phi]|^2 e^{-E_n/kT} \non \\
                   Z[U] &=& \int D\phi ~ e^{-H[U,\phi]/kT}  \ ,
\eea
where it is understood that $U,\phi$ on the left hand side of this equation are defined as the spatial links and scalar field on a three-dimensional time slice at $t=0$, and $kT$, in lattice units, is the inverse of the lattice extension in the time direction. We then ask whether the GCS ($Z_N$ for SU(N)) is spontaneously broken, for a given background $U$ in the system described by the partition function $Z[U]$.  For this purpose we introduce
\beq
          \ophi(x;U) = {1\over Z[U]} \int D\phi ~ \phi(x) e^{-H[U,\phi]/kT}  \ ,
\eeq
The global $Z_N$ symmetry is spontaneously broken
if $\ophi(x;U)$ is non-zero throughout the lattice.  In general this quantity is not the same at each position since the background $U$ breaks translation invariance, and typically the spatial average of  $\ophi(x;U)$ is negligible.  It is then convenient to define the spatial average of the modulus in the spatial volume $\V_3$ of the time slice
\beq
      \Phi[U] = {1\over \V_3} \sum_x |\ophi(x;U)| 
\label{PU}
\eeq
which is zero in the unbroken phase, and non-zero is the broken phase for a given background.  Having derived an order parameter for global symmetry breaking at a given $U$, we can now determine whether this symmetry is broken in the full theory by taking the
expectation value 
\beq
            \Phi \equiv \langle \Phi[U] \rangle  \ .
\eeq
in the usual $e^{-S}$ probability measure. If $\Phi > 0$, then the global center subgroup of the gauge group is broken in every relevant configuration generated by that measure, and this is
the precise meaning of the statement that global center symmetry is spontaneously broken in the Higgs phase.  
The order parameter $\Phi$ is very closely analogous to the Edwards-Anderson order parameter for a spin glass \cite{Edward_Anderson}, as emphasized
in \cite{Greensite:2020nhg}.
 
     Here we have glossed over an important technicality.   Strictly speaking, symmetries cannot be spontaneously broken in a finite
volume.  It is necessary to add a small term to $H[U,\phi]$, proportional to some constant $h$, which breaks the global symmetry explicitly.   Then the symmetry is broken if $\Phi > 0$ after taking first the thermodynamic and then the $h\ra 0$ limit.  It is possible to
construct this term in a way which does not break the local gauge symmetry.  For details the reader is referred to \cite{Greensite:2020nhg}; but for the numerical calculations discussed in the next section these formalities will not be necessary. 

\section{\label{line} Transition line in SU(2)$\times$ U(1) gauge Higgs theory}

    The electroweak gauge group has a special feature:  both the confined and the Higgs phases have charged states which couple
to a massless particle.\footnote{It is likely, however, that at sufficiently strong couplings there is a transition within the confinement phase to
a phase in which all vector particles are massive.  At least, this is known to be true in other gauge theories such as the abelian Higgs model, which also contains a U(1) gauge symmetry.}  An example of a state of this kind is the following:
\beq
  \Psi_{\text{chrg}} =  \pbar^a(x) \phi^a(x) \rho(x;V) \Psi_0 [U,V,\phi] \ ,
\eeq
while a neutral state can have the form
\beq
  \Psi_{\text{neutral}} =  \pbar^a(x) \sigma_1^{ab} \phi^{b*}(x) \Psi_0 [U,V,\phi] \ .
\eeq
and the $\s_i$ are the Pauli matrices.  Proposals of this kind (for electron and neutrino states) were made long ago in \cite{tHooft:1979yoe,Frohlich:1981yi,Banks:1979fi}.
Here $U,V$ are the SU(2) and U(1) lattice gauge fields respectively, and $\rho(x;V)$ is a pseudomatter field, analogous to \rf{rho}.
We take the static fermion operator $\psi$ and the Higgs field $\phi$ to both transform in the fundamental representation of SU(2),
but with opposite weak hypercharge $\pm \oh$. Thus  $ \Psi_{\text{chrg}}$ 
transforms under a global U(1) symmetry with weak hypercharge $-1$.   

In the Higgs phase, the global center symmetry $Z_2 \times U(1)$ is spontaneously broken.  Nevertheless, in the Higgs phase, an operator which transforms as a singlet under
global SU(2), and with hypercharge $-1$ under global U(1), can also be regarded as transforming covariantly under a certain global U(1) transformation which is unbroken in the Higgs phase.  The symmetry can be identified by going to a physical gauge defined by the
gauge rotating $\ophi$ such that upper component of $\ophi$ vanishes, i.e.
\beq
             \ophi^1(x;U) = 0 \ ,
\label{gcond}
\eeq
and in this gauge there is a remnant local U(1) symmetry consisting of transformations
\beq
           g(x) = e^{i\a(x) \s_3/2} e^{i\a(x)/2} \ .
\eeq      
Under a global transformation $\a(x)=\a$ of this type, $\Psi_{\text{chrg}}  \ra e^{i\a}  \Psi_{\text{chrg}}$ in the Higgs phase.  In the confined phase, where
the global GCS is unbroken and the gauge \rf{gcond} is ill-defined, we still have $ \Psi_{\text{chrg}}  \ra e^{i\a}  \Psi_{\text{chrg}}$ under a $Z_2 \times U(1)$ transformation
$g(x) = (\pm \mathbb{1})e^{i\a/2}$.  

    The procedure for finding the confinement to Higgs transition in SU(2) gauge theories has been described in \cite{Greensite:2020nhg}, and this procedure
 is unchanged for SU(2)$\times$U(1).  The lattice action is
 \bea
 S  &=& -\b \sum_{plaq} \left[\oh \tr[UUU^\dg U^\dg] + {1\over \tan^2(\th_W)} \text{Re}[VVV^\dg V^\dg] \right] \non \\
      & &  -\g \sum_{x,\m} \text{Re}[\phi^\dg(x) U_\m(x) V_\m(x) \phi(x+\hat{\m})] \ ,
 \eea
 with SU(2) gauge field $U_\m(x)$ and U(1) gauge field $V_\m(x) = e^{i\th_\m(x)}$ and,
 for simplicity, we have imposed the unimodular condition $|\phi|=1$.  For the lattice versions of the $W, Z$ and photon fields in
 terms of $U,V$ in unitary gauge,  cf.\ Veselov and Zubkov \cite{Zubkov:2008gi}.
The gauge and scalar fields are updated in the usual way, but each data-taking sweep actually consists of a set of $n_{sym}$ sweeps in which the spacelike links $U_i(\vx,0),V(\vx,0)$ are held fixed on the $t=0$ time slice.  Let $\phi(\vx,t=0,n)$ be the scalar field at site $\vx$ on the $t=0$ time slice at the $n$-th sweep.  Then we compute $\ophi(\vx,U,V)$ from the average over $n_{sym}$ sweeps 
\beq
             \ophi(\vx,U,V) = {1\over n_{sym}} \sum_{n=1}^{n_{sym}} \phi(\vx,0,n) \ ,
\eeq
and  the order parameter $\Phi_{n_{sym}}(U,V)$ from \rf{PU}.  Here it is important to indicate the dependence on $n_{sym}$.
Then the procedure is repeated, updating links and the scalar field together, followed by another computation of 
$\Phi(n_{sym},U)$ from a simulation with spatial links at $t=0$ held fixed, and so on.    Averaging the $\Phi(n_{sym},U)$ obtained by these means results in an estimate for $\langle  \Phi_{n_{sym}}(U) \rangle$.  Since $\Phi_{n_{sym}}(U)$ is a sum
of moduli, it cannot be zero.  Instead, on general statistical grounds, we expect~\footnote{One must keep in mind that 
at finite $\V_3$, $ \langle \Phi \rangle$ would actually vanish at ${n_{sym} \ra \infty}$, since a symmetry cannot break in a finite volume.  The proper order of limits is first $\V_3 \ra \infty$, then $n_{sym} \ra \infty$.  Nevertheless, for $n_{sym}$ not too large,  \rf{fit} is a good fit to the data, and the extrapolation should be reliable.} 
\beq
           \langle \Phi_{n_{sym}}(U,V) \rangle = \Phi + {\k \over \sqrt{n_{sym}}} \ ,
\label{fit}
\eeq
where $\k$ is some constant.  
By computing  $\langle  \Phi_{n_{sym}}(U,V) \rangle$ in independent runs at a range of $n_{sym}$ values, and fitting
the results to \rf{fit}, we obtain an estimate for $\Phi$ at any point in the  $\b,\g$ plane of lattice couplings. 

\subsection{Numerical results}

   We work throughout at fixed Weinberg angle $\th_W=0.5002$ radians. In this section we allow both $\b$ and $\g$ to vary 
 and use the approach just described to compute the confinement to Higgs transition line in the $\b-\g$ plane. It is understood
 that in both phases there exists a massless vector boson, so not all charges are confined in either phase.   But in the confined phase, at least the non-abelian charge is confined according to the \Sc criterion.  In the next section, for the computation of vector boson masses,
 we set $\b=10.1$, which should be close to the physical value corresponding to the usual fine structure constant. 
 
\begin{figure}[htb]
\centerline{\includegraphics[scale=0.6]{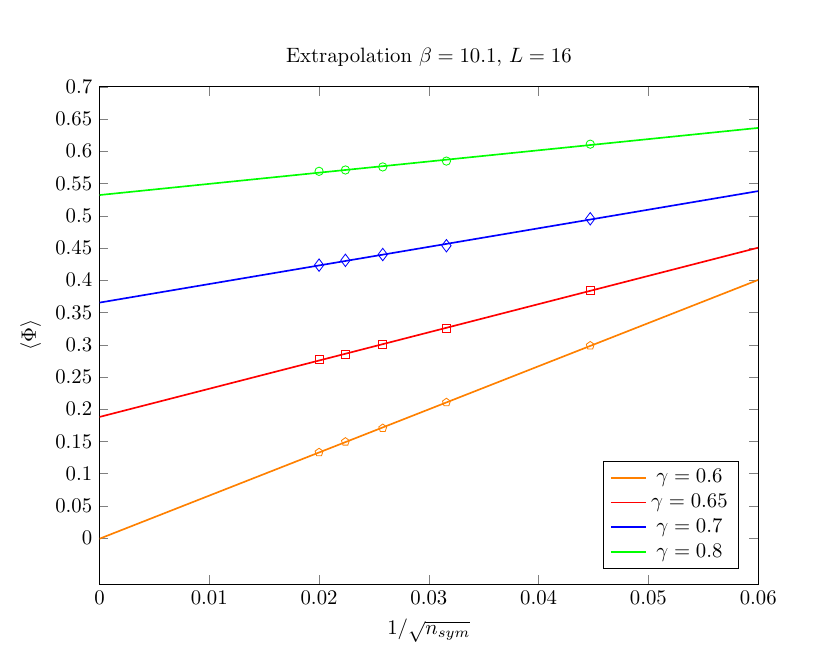}}
 \caption{Computation of the order parameter $\langle \Phi_{n_{sym}}\rangle$ vs.\ $1/\sqrt{n_{sym}}$ where $n_{sym}$ is number of Monte Carlo sweeps (see text).  Also shown is the extrapolation to $n_{sym}=\infty$.  Data was taken at  $\b=10.1$ for the set of $\g$ shown.  
 Extrapolation to $\langle \Phi \rangle =0$ indicates that the system is in the \Sc confined phase, while extrapolation to 
 $\langle \Phi \rangle > 0$ means that the system is in the Higgs phase.}
  \label{nsym}
\end{figure}

\begin{figure}[htb]
 \includegraphics[scale=0.55]{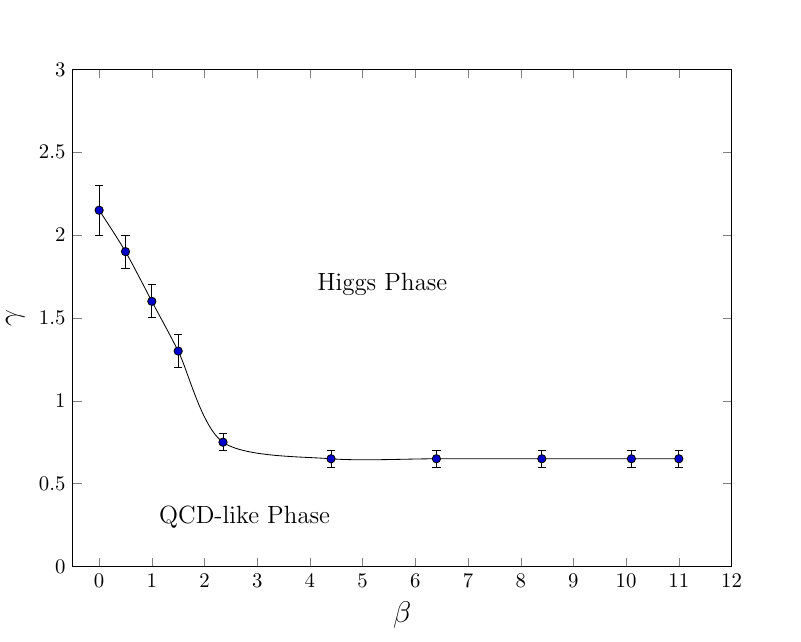}
 \caption{Transition line from the symmetric to the broken phase of global center symmetry, corresponding to the transition from
 the confinement to the Higgs phase, in the $\b-\g$ plane at fixed $\th_W$. }
 \label{phase}
\end{figure}

    In Fig.\ \ref{nsym} we display the extrapolation of $\langle \Phi_{n_{sym}}(U,V) \rangle$ to $n_{sym} \ra \infty$ at $\b=10.1$ on a $16^4$
lattice volume.  For $\g \le 0.6$, the data extrapolates to $\Phi=0$, and the system is in the confined phase.  At $\g=0.65$ and above, 
$\Phi>0$, and the system is in the Higgs phase.  The transition is for $\g$ somewhere between 0.6 and 0.65.  Fig.\ \ref{phase} shows the transition line obtained by this method, up to $\b=11$, on $\V=16^4$ lattice volumes.\footnote{It is possible that the sudden change in slope of the transition line around $\b=2$ is associated with a transition to a fully massive phase in the region of unbroken GCS.  A similar effect was seen in SU(2) gauge theory in five dimensions, cf.\ Ward  \cite{Ward:2021qqh}.}   

 As noted in the Introduction, the first results for
the SU(2)$\times$U(1) phase diagram with a fixed modulus Higgs were obtained by Shrock \cite{Shrock:1985un,Shrock:1985ur,Shrock:1986av},  who in fact presented
phase boundary surfaces in the full three dimensional phase volume (two gauge couplings and $\g$).   In this early work the criterion for a transition was thermodynamic, i.e.\ non-analytic behavior of the plaquette and the Higgs action density $\varphi$ defined below, and results were obtained by a combination of series expansions around soluble limits, and the numerical tools and methods available in the mid 1980s  (the latter included estimation of transition points from hysteresis curves on very small lattices).   We are calculating essentially a slice of the phase diagram in the three volume, and although it is difficult to make a precise comparison, our transition line, derived using a different criterion for the Higgs phase, appears to roughly agree with the results reported by Shrock.   Our transition line is also very similar, but perhaps not exactly the same, as the line obtained in much later work by Veselov and Zubkov \cite{Zubkov:2008gi}.  Those authors, however, used slightly different parameters for the (renormalized) fine structure constant and the Weinberg angle, so some modest deviation is to be expected. Their criterion for a transition is also different from ours, and has to do with a drop in monopole density.

\begin{figure}[htb]
 \includegraphics[scale=0.5]{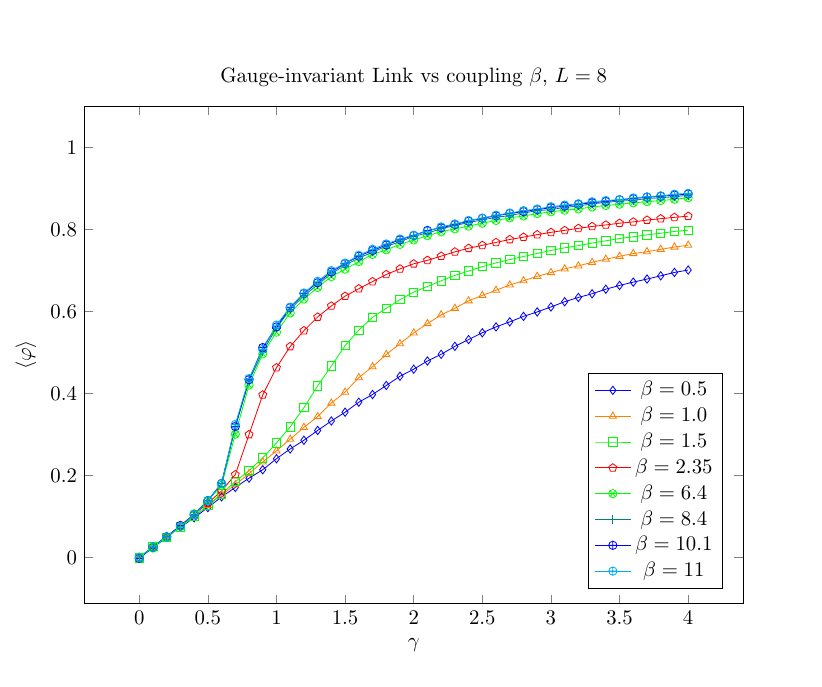}
 \caption{Expectation value $\langle \varphi \rangle$ of the gauge invariant link \rf{vphi} vs.\ $\g$ at several $\b$ values
 on an $8^4$ lattice volume.}
 \label{linksv8}
\end{figure}

\begin{figure}[htb]
 \includegraphics[scale=0.5]{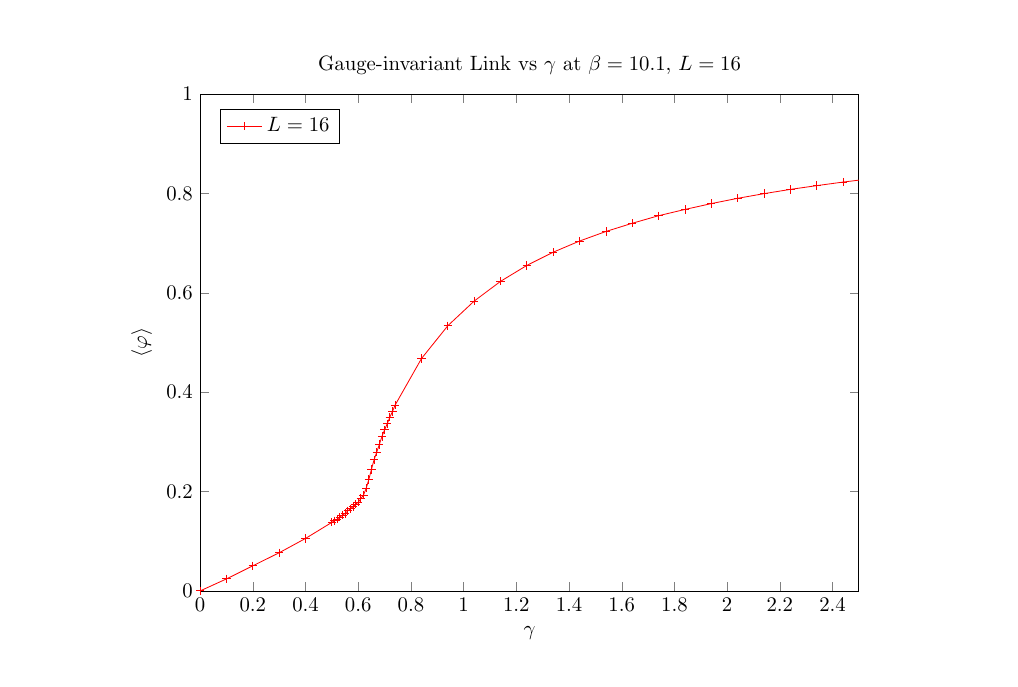}
 \caption{Same as Fig.\ \ref{linksv8} for $\b=10.1$ on a $16^4$ lattice volume. }
 \label{Link16}
\end{figure}

Since our criterion for a transition to the Higgs phase is non-thermodynamic, and the transition may or may not be accompanied by a thermodynamic transition, the next question is whether the confinement to Higgs transition line is also a line of thermodynamic
transition.  The answer appears to be similar to the SU(2) case: the symmetry breaking transition is a thermodynamic transition
at large $\b$, but not at small $\b$.   Define the ``gauge-invariant link'' as the spacetime average of the Higgs action
\beq
 \varphi = {1\over 4\V_4} \sum_{x,\m} \text{Re}[\phi^\dg(x) U_\m(x) V_\m(x) \phi(x+\hat{\m})]
\label{vphi}
\eeq
where $\V_4$ is the lattice 4-volume. In Fig.\ \ref{linksv8} we plot the expectation value of $\varphi$ vs.\ $\g$ at various $\b$ on a $\V_4=8^4$ volume, and we see that while the curve is smooth at small $\b$, it seems to develop a ``kink,'' i.e.\ a discontinuity in the slope $\pa \varphi / \pa \g$ at larger $\b$.  In Fig.\ \ref{Link16} we show the same data for $\b=10.1$ on a $16^4$ lattice volume.  Precisely this type of 
non-analyticity has been seen before in the transition to the Higgs phase in the abelian Higgs model \cite{Matsuyama:2019lei}.  It is also useful to study the
susceptibility, 
\bea
           \chi_\varphi &=&  4\V_4(\langle \varphi^2\rangle - \langle \varphi\rangle^2)  
\label{suscept}
\eea
at $\b=10.1$, vs.\ $\g$ at various volumes.  Note that the data point for the $12^4$ volume at the transition point lies well above the data
points at lower volumes.  All of this suggests a thermodynamic transition or, at least, a very sharp crossover around $\g=0.62$ at $\b=10.1$, which is consistent with our estimate of the location of the GCS breaking transition, somewhere between $\g=0.60$ and $\g=0.65$.

\begin{figure}[htb]
 \includegraphics[scale=0.6]{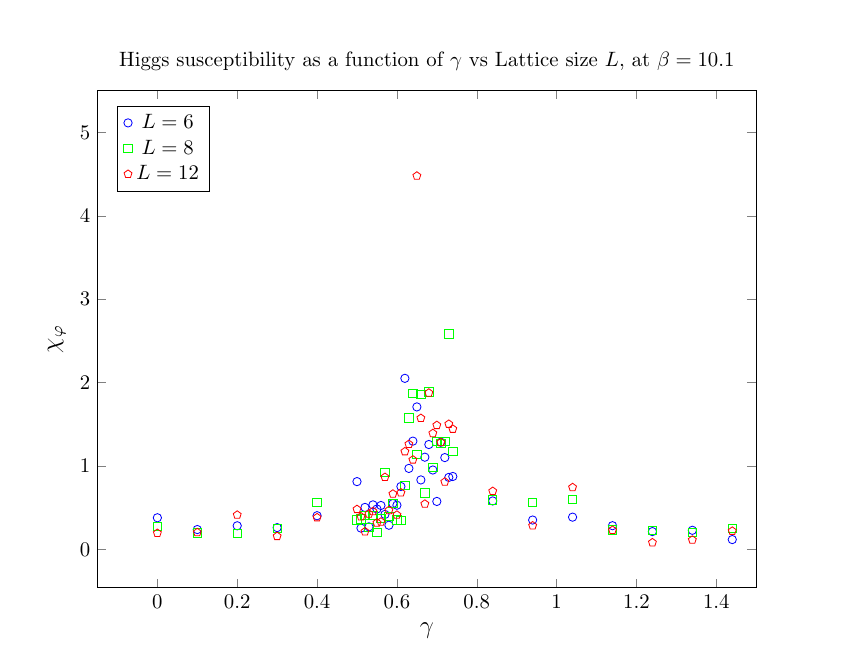}
 \caption{Link susceptibility \rf{suscept} vs.\ $\g$ at $\b=10.1$ and various lattice volumes $\V=L^4$.}
 \label{Sus}
\end{figure}

\section{\label{newZ}New weak vector bosons?}

     We search for neutral vector bosons in our simplified SU(2)$\times$U(1) gauge theory by studying correlation functions
of gauge invariant operators which, operating on the vacuum, would create
physical states corresponding to neutral vector bosons.  Photons and physical Z bosons are states of this type, and we would like to know if there are any more in the spectrum.  Define $\tilde{U}_\m(x) = U_\m(x) V_\m(x)$, with the lattice Laplacian operator $D^{ab}_{xy}[\tilde{U}]$ covariant under the SU(2)$\times$U(1) group.  Let $\zeta_i(x), i=1,2..,N$ 
be the lowest $N$ eigenstates of the Laplacian operator, with 
$\zeta_{N+1}(x)=\phi(x)$ the Higgs field.

Now we use the $\zeta$ fields to construct gauge-invariant operators which, when operating on the vacuum,
will construct physical states with the quantum numbers of the physical photon and Z.  Define
\bea
    \rho(\vx) e^{i{\cal A}^i_\m(\vx)} = \zeta_i^\dg(x) \tilde{U}_\m(\vx,t) \zeta_i(\vx+\hat{\m})  \ ,
\eea
and
\bea
     A^i_\m(\vx) &=& \sin({\cal A}^i_\m(\vx)) \non \\
     Q_\m^i &=& {1\over L^3} \sum_\vx A^i_\m(\vx) \non \\
     |\Phi_\m^i\rangle &=& Q^i_\m |\Psi_0\rangle \ ,
\label{Q}
\eea
where index $\m=1,2,3$ are spatial directions, and $i$ (which is labels the choice of pseudomatter (or Higgs) field 
$\zeta_i$.   
 
 Let $T=e^{-Ha}$  be the transfer matrix, $\E$ 
the vacuum energy, and
\beq
            \t = T e^{\E a} = e^{-(H-\E)a}
\eeq
a modified transfer matrix.  Consider two states $|A\rangle = A|\Psi_0\rangle$ and $|B\rangle = B|\Psi_0\rangle$.  Then 
\beq
           \langle A| \t^t |B \rangle \equiv \langle A| e^{-(H-\E)t}|B \rangle = \langle A^\dg(t) B(0) \rangle \ .
\eeq
 
 We look for states $\Psi_\m^n$ which diagonalize $\t$ in the subspace of Hilbert space spanned by the $|\Phi_\m^i\rangle$ such that
\bea
            \langle \Psi^m_\m| \Psi^n_\m \rangle &=& \d_{mn} \non \\
             \langle \Psi^m_\m|\t| \Psi^n_\m \rangle &=& \l_n \d_{mn} \ .
\eea
To achieve this, we compute the ${(N+1)\times (N+1)}$ matrices
\bea
            O_{ab} &=&  \langle\Phi_\m^a|\Phi_\m^b\rangle \non \\
            T_{ab}  &=& \langle \Phi_\m^a|\t|\Phi_\m^b\rangle \ .
 \eea
 Then we solve numerically the generalized eigenvalue equation
 \beq
              T_{ab} v^n_b = \l_n O_{ab} v^n_b \ ,
 \eeq
 or
 \beq
            [T] \vec{v}^n = \l_n [O] \vec{v}^n \ .
 \eeq
 There will be $N+1$ vectors $\vec{v}^n$ which satisfy this equation, and then
 \beq
              |\Psi^n_\m\rangle = \sum_a v^n_a |\Phi_\m^a\rangle 
 \eeq
 are the eigenstates of $\t$ in the subspace.   Now we evolve these states in Euclidean time, and define
 \bea
             G_n(t) &=& \langle \Psi_\m^n| \t^t |\Psi_\m^n\rangle \non \\
                        &=& \sum v^{n*}_a v^n_b \langle \Phi_\m^a | \t^t | \Phi_\m^b \rangle  \ .
 \eea
 Since, on general grounds,
 \beq
    |\Psi_\m^n \rangle = \sum_i  c^{ni}_\m |i\rangle \ ,
 \eeq
 where the $|i\rangle$ are energy eigenstates (i.e.\ exact eigenstates of the Hamiltonian),
 it follows that
 \beq
            G_n(t) = \sum_i  |c^{ni}_\m|^2 e^{-E_i t} \ ,
\eeq 
where $E_i$ is the energy of state $|i\rangle$ above the vacuum energy (i.e.\ it is the energy minus $\E$).

     By construction all states are zero momentum, so for the one-particle states these
are particles at rest.  In that case, their masses correspond to the $E_i$.   If each of the $\Psi_\m^n$ were exact eigenstates
of the transfer matrix in the full Hilbert space, then
\beq
          G_n(t) = a e^{-E_q t} \ ,
\eeq
where $E_q$ is one of the energy eigenvalues; in our case one of the particle masses.  But that seems unlikely; we do not expect
that the $\{\Psi_\m^n\}$ are exact eigenstates of the Hamiltonian.  In general it is difficult to fit data to
a sum of exponentials, unless the data is extraordinarily accurate.  However, in our case we actually know for sure one
of the masses, which is the mass of the photon, and that mass is zero.  If we are fortunate, it may be enough fit each of the $G_n(t)$ to
the simple form
\beq
          G_n(t) = a e^{-bt}  + c \ ,
\label{form}
\eeq
where $b$ is a non-zero particle mass, and $c = c e^{-0t}$ is coming from an admixture of the massless photon state. 

Numerically we compute
\beq
           \langle \Phi_\m^j| \t^t | \Phi_\m^i \rangle =  \langle Q^{j\dg}_\m(t) Q^i_\m(0) \rangle
\label{corr}
\eeq
by lattice Monte Carlo.
Note that $t=0$ corresponds to matrix $O_{ab}$ and $t=1$ gives us $T_{ab}$.  This provides the necessary information
to determine the correlators $G_n(t)$ described above.  
 
\subsection{Numerical results}

We work on a $16^3 \times 36$ lattice at $\th_W=0.5002$ radians and $\b=10.1$.  At tree-level, as in the continuum theory, there
are two neutral vector bosons: a massless photon, and the Z boson. Setting the physical value $m_Z^{phys} = m_Z/a = 91.2$ GeV determines the lattice spacing, and hence $\g$ in physical units
\beq
\sqrt{\g^{phys}} = {\sqrt{\g} \over m_Z} ~ 91.2 ~\text{GeV}
\eeq
In the electroweak sector of the Standard Model, with a finite Higgs mass and dynamical fermions, the known value
is $\g^{phys}=246$ GeV.  The Z mass at tree level is 
\beq
           m^{pert}_Z = \sqrt{\g \over \b} {1\over \cos \th_W}
\label{mtree}
\eeq

\begin{figure}[htb]
\subfigure[~$\g=2$]  
{ 
 \includegraphics[scale=0.55]{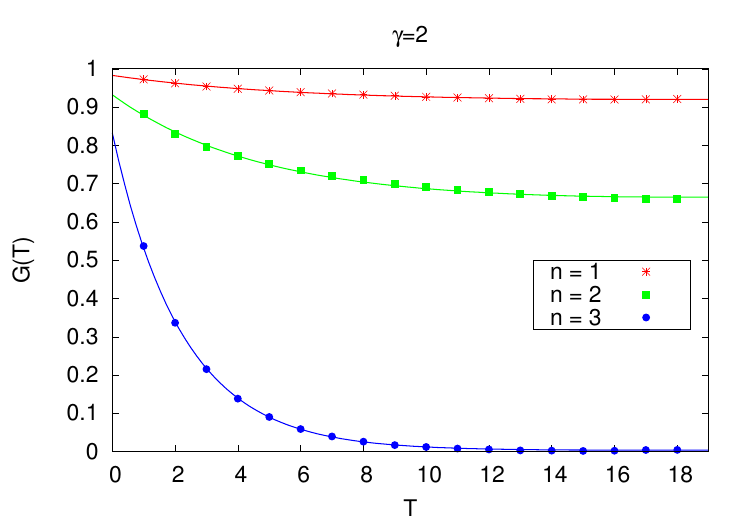}
 } 
\subfigure[~$\g=4$]
{ 
 \includegraphics[scale=0.55]{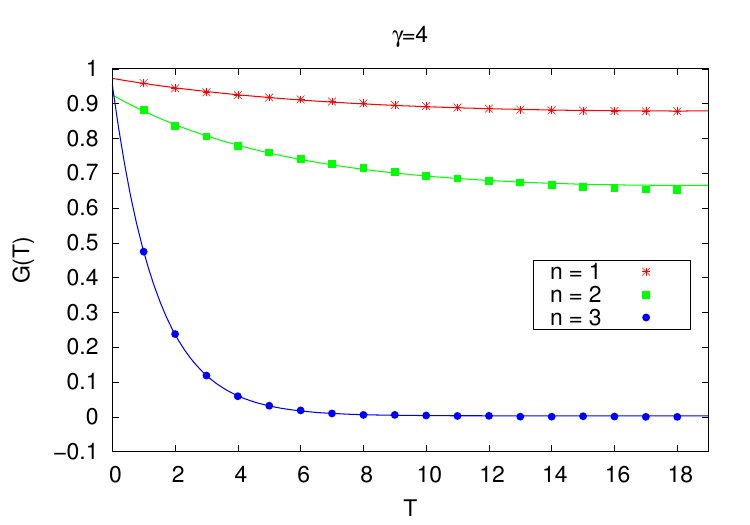}
}
\subfigure[~$\g=6$]
{ 
 \includegraphics[scale=0.55]{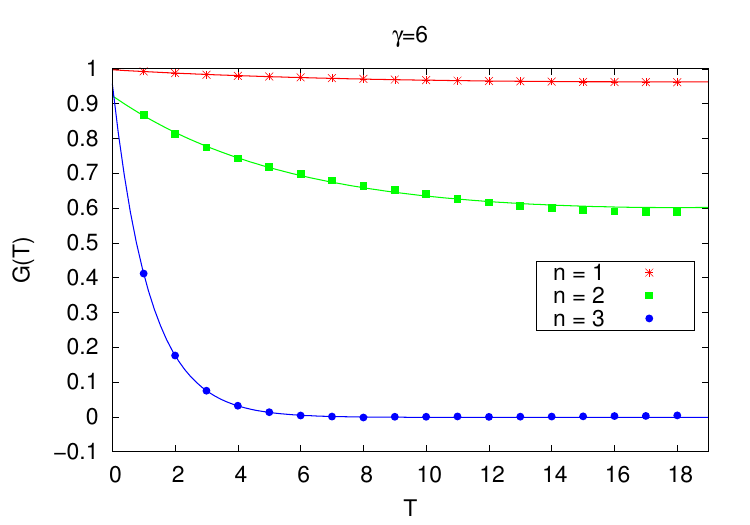}
 }
 \subfigure[~$\g=8$]
{ 
 \includegraphics[scale=0.55]{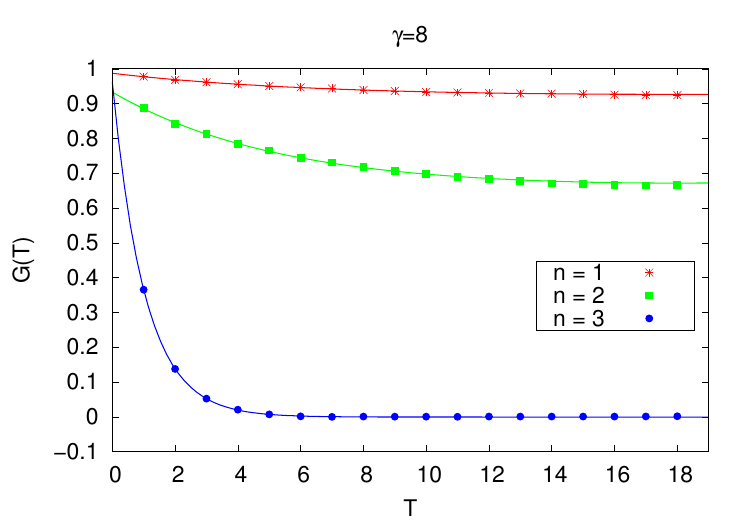}
 }
 \caption{$G_n(t)$ vs.\ $t$ at $\g=2,4,6,8$ and $\b=10.1$.}
 \label{Gall}
 \end{figure}

In Fig.\ \ref{Gall} we show our data for $G_n(t)$ at $n=1,2,3$ and $\g=2,4,6,8$, $\b=10.1$, obtained using $N=3$ Laplacian eigenstates and the Higgs field as described above.  It is clear that the state $\Psi_\m^1$ is mostly the massless
photon state, with only a small admixture of higher energy densities, given the fact that $G_1(t)$ asymptotes to a flat line with $G_1(t) > 0.9$.
Because the falloff over a range of $1\le t\le 18$ is so small, it is difficult for us to reliably extract the admixture of higher mass.  So this mainly photon state will be
excluded in the following logarithmic plots.  The $n=2$ state also has a substantial admixture of a photon state, but for $n=2$ and $n=3$
the situation is more favorable for extracting masses, as we will see.  Since our truncated Hilbert space is spanned by four states,  there
is also data for $n=4$, but here the data is rather noisy, and not favorable for curve fitting.

\clearpage

At each $\g$ value we have run 20 independent lattice Monte Carlo simulations consisting of 250,000 sweeps each, with 20,000 sweeps for thermalization, and data taking separated by 100 update sweeps.  Each independent run supplies data for 
\beq
\langle \Phi_\m^a | \t^t | \Phi_\m^b \rangle
\eeq
with $a,b =1,2,3,4$.  The linear algebra required to compute $G_n(t)$ is carried out by a Matlab program.  To plot the data we simply average the 20 values of $G_n(t)$ at each $t$, and compute $\Delta G(t)$ from the standard deviation. Masses can be extracted from a fit to
\beq
           G_n(t) = a ( e^{-m_n t} + e^{-m_n(36-t)}) + c
\label{Gn}
\eeq
where we have allowed for periodicity in the time direction.  However, $\Delta G_n(t)$ cannot be interpreted as an error bar, because the chi-square of the curve fit is far too low.  It is better to think of these as indicating an envelope which contains 20 separate smooth curves.  An alternative procedure is to fit each of the twenty data sets to the form \rf{Gn}, resulting in 20 values for $m_n$.  This gives us an average and
an error bar for the $m_n$, which are the values shown in Table \ref{masses-errors}.  The masses $m_n$ are in good agreement with the masses obtained by the first method, although we regard the second method as the appropriate procedure for obtaining error estimates
on the mass values.

\begin{figure}[htb]
\subfigure[~]  
{
 \includegraphics[scale=0.55]{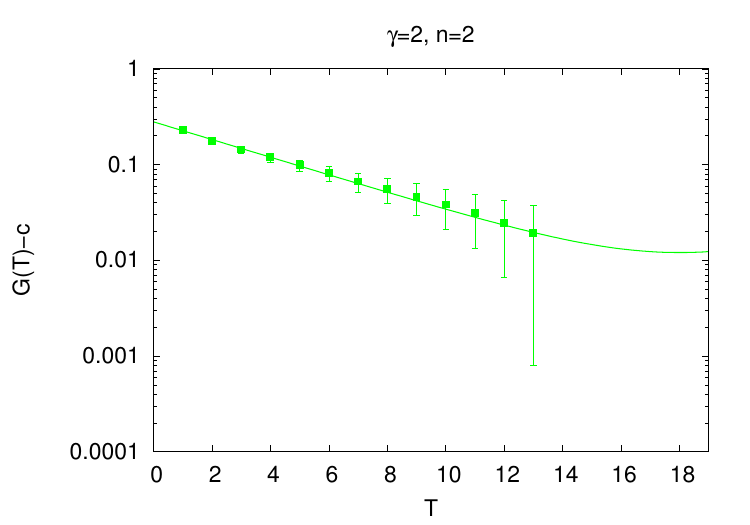}
 } 
\subfigure[~]
{
 \includegraphics[scale=0.55]{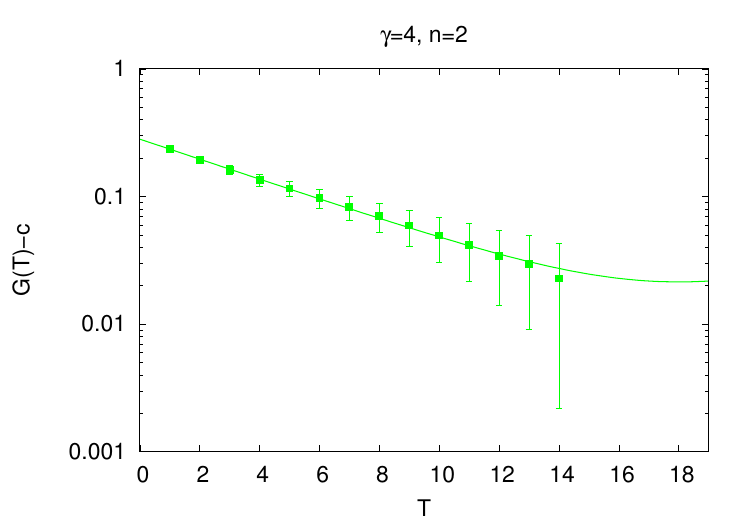}
}
\subfigure[~]
{
 \includegraphics[scale=0.55]{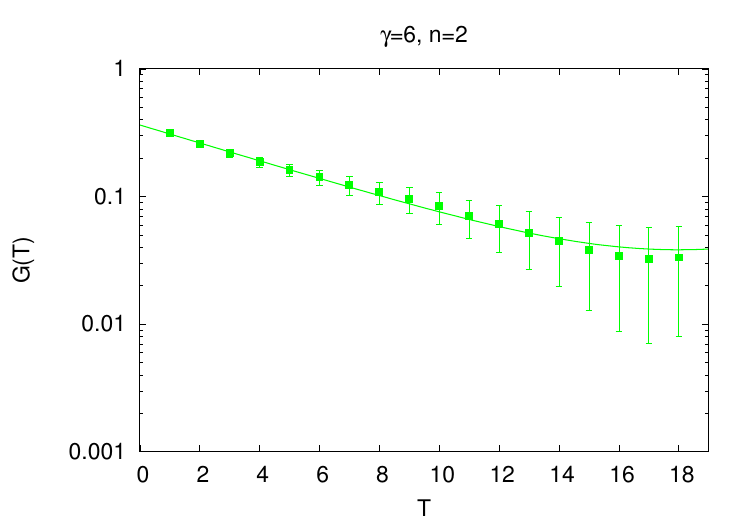}
 }
 \subfigure[~]
{
 \includegraphics[scale=0.55]{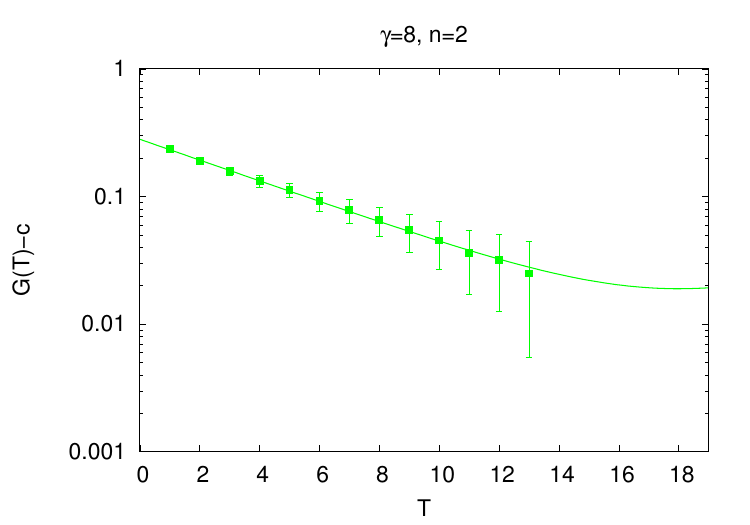}
 }
 \caption{Logarithmic plot of $G_2(t)$ with the photon contribution (the constant $c$) subtracted from the data and the fitting
  function, at \\ $\g=2,4,6,8$.}
 \label{n2}
 \end{figure}

\begin{figure}[htb]
\subfigure[~]  
{
 \includegraphics[scale=0.55]{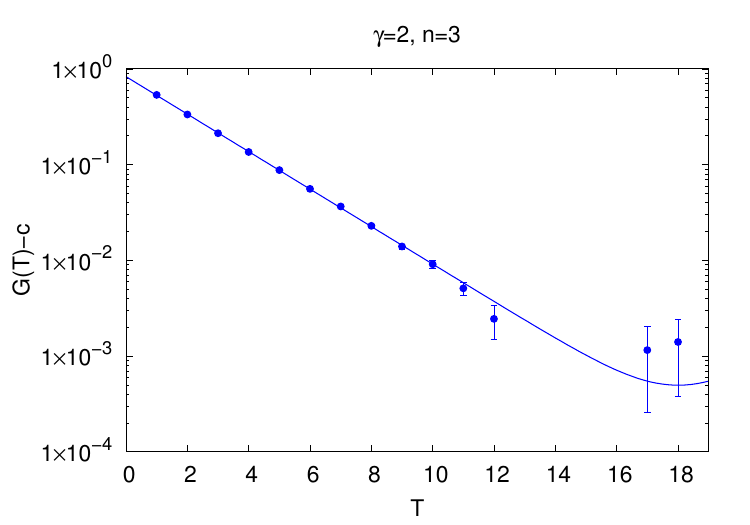}
 } 
\subfigure[~]
{
 \includegraphics[scale=0.55]{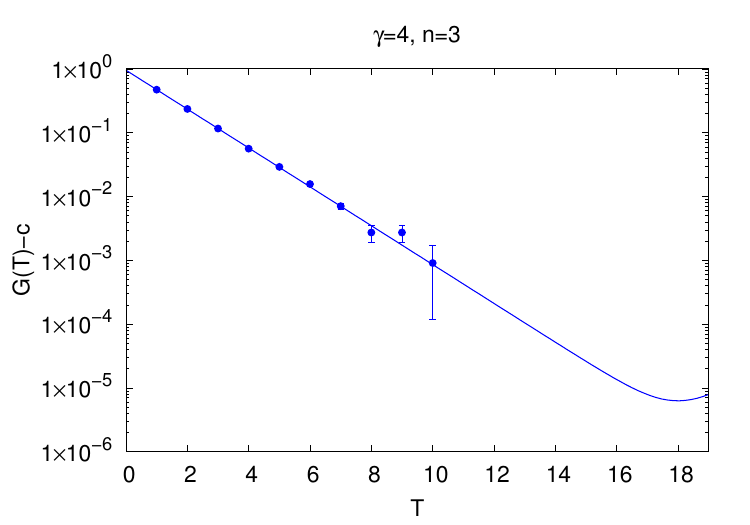}
}
\subfigure[~]
{
 \includegraphics[scale=0.55]{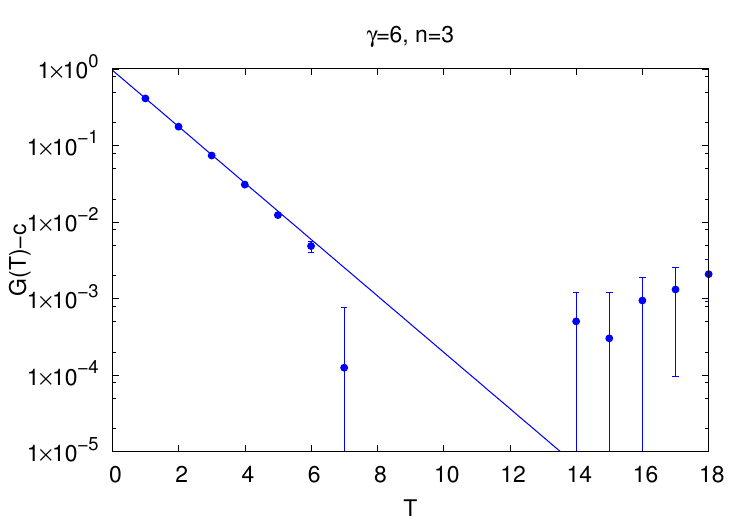}
 }
 \subfigure[~]
 
{
 \includegraphics[scale=0.55]{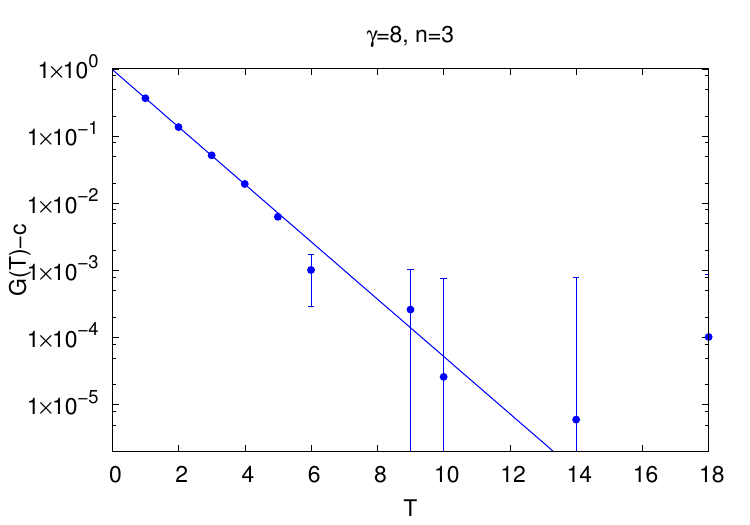}
 }
 \caption{Same as the previous figure, but for $G_3(t)$.}
 \label{n3}
 \end{figure}

   In Figs.\ \ref{n2} and \ref{n3} we plot the data for $G_2(t)-c$ and $G_3(t)-c$ respectively with the fitting constant $c$ subtracted from both the data.  Having subtracted the photon component represented by the constant $c$ from the data, we display the subtracted data compared to the previous fit also with $c$ subtracted, i.e.
\beq
a ( e^{-m_n t} + e^{-m_n(36-t)})
\eeq
The data appears to fit a straight line on a log plot, at least for the smaller $T$ values.  Deviations at larger $T$ values may be attributed to the fact that when the fitting constant $c$ is quite small, which is generally the case for $n=3$, then small fluctuations in the $G_n(t)$ data result in seemingly large deviations in 
$G_n(t)-c$ from a straight line fit on a logarithmic plot.
 
\begin{figure}[htb]
 \includegraphics[scale=0.55]{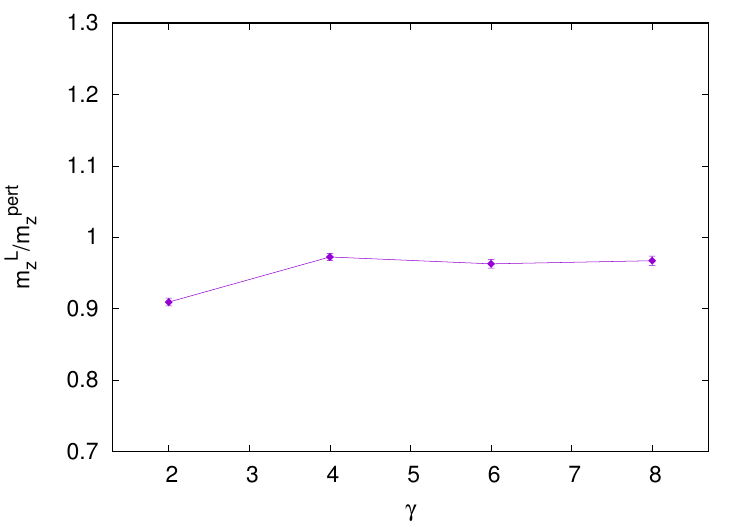}
 \caption{The ratio of the computed $m_3$ (denoted $m_Z^L$) mass obtained from lattice simulations to the tree-level value $m^{pert}_Z$ of the Z-boson mass (eq.\ \rf{mtree}).}
 \label{mz}
 \end{figure}

\begin{figure}[htb]
 \includegraphics[scale=0.55]{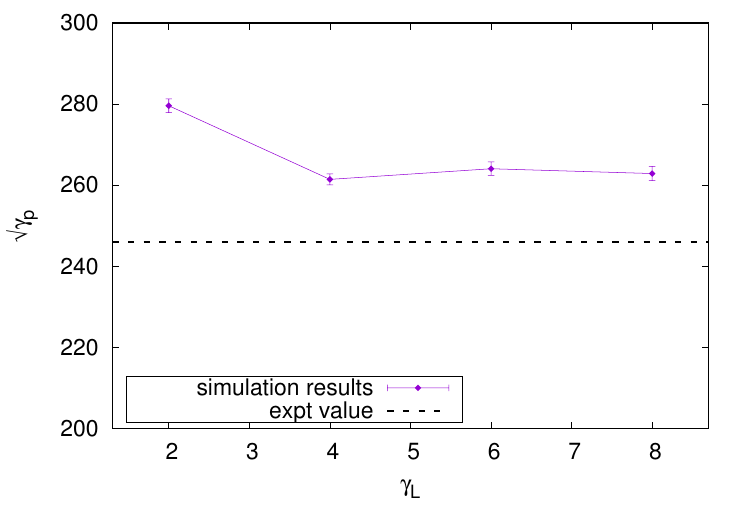} 
 \caption{From the identification of $m_3=m_Z$ and the known mass of 91.2 GeV of the Z, we obtain, for each lattice $\g$, the
 value of $\sqrt{\g}$ in physical units.  The accepted value in the electroweak theory is $\sqrt{\g^{phys}}\approx 246$ GeV, shown as a dashed line. }
 \label{gamma}
 \end{figure}

    In Fig.\ \ref{mz} we show the ratio of the mass of the $n=3$ state, determined from a fit to the $G_3(t)$ data, divided by the lattice tree level value $m_Z^{pert}$ in \rf{mtree}, and we see that the ratio, especially for $\g>2$, is quite close to unity, which gives us confidence
in identifying the $n=3$ mass as the mass of the Z boson obtained in our simulations.   Using these values for $m_Z$, we obtain a lattice spacing and a corresponding value for $\sqrt{\g^{phys}}$ for each lattice coupling $\g$.  These results (Fig.\ \ref{gamma}) come out close to the experimental value of $\sqrt{\g^{phys}} \approx 246$ GeV.  All of this, together with the appearance of a zero mass (photon) state in the data, gives us some confidence that our procedure is delivering the results expected from perturbation theory.

 \clearpage          
           
  \begin{figure}[htb]
\label{Eg1p2}
 \includegraphics[scale=0.55]{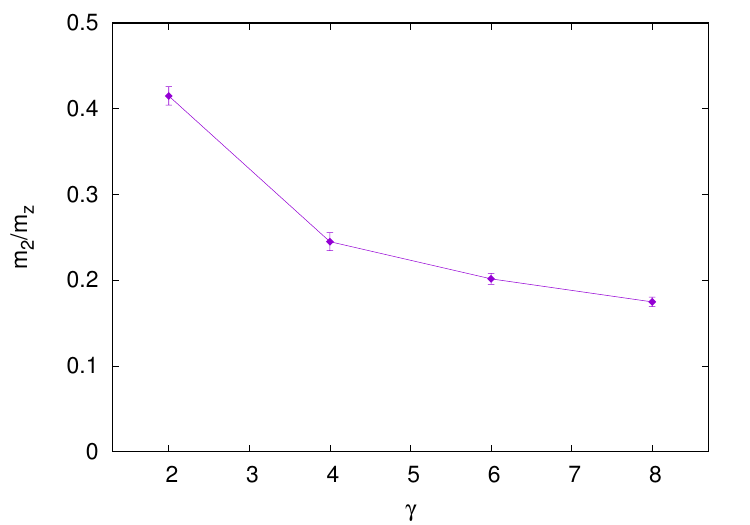}
 \caption{Ratio of $m_2/m_3$, where we have identified $m_3=m_Z$.}
 \label{nonpert}
 \end{figure}    
 
 \clearpage
 
   But what do we make of the intermediate mass $m_2$?  The data supporting the existence of such a state, shown in Fig.\ \ref{n2},
seems just as solid as the data for $m_Z=m_3$, and if we take this data seriously, it appears that the lattice regularized SU(2)$\times$U(1)
theory contains an extra vector boson state, in addition to the photon and Z boson states that we have already seen, that is
invisible in perturbation theory.  The problem, however, is that $m_2$ in lattice units is almost insensitive to $\g$, and as a result the
mass ratio $m_2/m_Z$ is $\g$-dependent, as seen in Fig.\ \ref{nonpert}.  A first thought, since this is by construction a zero total momentum state, is that $m_2$ might represent two photons of opposite momenta.  But for a lattice of 16 units in spatial extent, this would be a state of
energy $\ge$ 0.79 in lattice units, which is far above $m_2$.   Moreover, we have checked our results on a smaller lattice of volume
$12^3 \times 36$, with results for masses consistent (to within a few percent) with the results shown in Table \ref{masses-errors}.  It seems that the data insists that $m_2$ is the mass of a static massive one-particle state. 
But this implies non-universality of the lattice action, at least so far as this intermediate mass particle is concerned.   As for reasons, we can only speculate that it might be related to our unimodular condition
on the Higgs field, or to the triviality of lattice $\phi^4$ theories in general.

\begin{table}[ht]
    \centering
    \begin{tabular}{|c|c|c|}
    \hline
    \multirow{3}{*}{$\gamma$} & \multicolumn{2}{c|}{mass in lattice units} \\
    \cline{2-3}
                              & intermediate boson $n_2$& Z boson ($n_3$)\\
    \hline
    2                         & $0.1914 \pm 0.0048$& $0.4612 \pm 0.0028$\\
    \hline
    4                         & $0.1709 \pm 0.0072$& $0.6975 \pm 0.0036$\\
    \hline
    6                         & $0.1705 \pm 0.0052$& $0.8458 \pm 0.0053$\\
    \hline
    8                         & $0.1714 \pm 0.0053$& $0.9810 \pm 0.0066$\\
    \hline
  \end{tabular}
    \caption{Masses in lattice units of the intermediate mass vector boson, extracted from $G_2(t)$, and the Z boson, extracted from $G_3(t)$, for several values of $\gamma$ at $\b=10.1$.}
  \label{masses-errors}
\end{table}

\section{\label{conclude}Conclusions}
 
 In this article we have considered an SU(2)$\times$U(1) gauge Higgs theory with a unimodular Higgs field, fixed Weinberg angle,
 and no dynamical fermions.  In this simplified theory we have located the transition line between the confinement and Higgs phases,
 as determined from the spontaneous breaking of the global center symmetry of the gauge group.  Then, in the Higgs phase, we have found at each parameter $\g$, with fixed Weinberg angle and fine structure constant, a set of three neutral vector bosons.  One of these is the massless photon, and a second can be identified, because of the proximity of its mass to the tree level Z boson mass, as the Z boson.  But we have also
found another massive particle state, well below the mass of the Z, which seems to be entirely non-perturbative in origin.  However, the ratio of
mass of this ``light Z'' particle to the Z mass varies with $\g$, which indicates a non-universality in the lattice formulation.   If we had obtained
results in which this ratio were fixed, then we would be able to quote a value of $m_2$ in physical units, and go on to wonder why such a
state has not (yet?) been seen in the collider data.  However, in view of the non-universality of this ratio, such phenomenological considerations
seem premature.  We speculate
that this non-universality might be associated with either our unimodular constraint on the Higgs field, or perhaps the triviality of $\phi^4$ theories.  In any case, because of this non-universality, we cannot offer any predictions about the existence of a light Z in the physical spectrum of the electroweak theory.   It would be interesting to repeat the calculation with a realistic Higgs potential, and we hope to report
on the results in a future publication. \\

 \acknowledgments{This research is supported by the U.S.\ Department of Energy under Grant No.\ DE-SC0013682.}   

 \bibliography{sym3}

 \end{document}